\documentclass[aps,pra,reprint,superscriptaddress,letterpaper]{revtex4-1}

\usepackage{amsmath,amsthm,amsfonts,amssymb,bm}
\usepackage{times}
\usepackage[colorlinks={true},dvipdfm]{hyperref}
\hypersetup{citecolor={blue}, filecolor={blue}, linkcolor={blue}, urlcolor={blue}}
\usepackage{float}
\usepackage{epsf}
\usepackage{color}
\usepackage{verbatim}
\usepackage{enumerate}
\usepackage{multirow}
\usepackage{anysize}
\usepackage{graphicx}
\usepackage{comment}
\usepackage{bm}

\begin{document}
\title{Searching for an optimal control in the presence of saddles on the quantum mechanical observable landscape}
\author{Gregory Riviello}
\affiliation{Department of Chemistry, Princeton University, Princeton, New Jersey 08544, USA}
\author{Re-Bing Wu}
\affiliation{Department of Automation, Tsinghua University and Center for Quantum Information Science and Technology, TNlist, Beijing, 100084, China}
\author{Qiuyang Sun}
\affiliation{Department of Chemistry, Princeton University, Princeton, New Jersey 08544, USA}
\author{Herschel Rabitz}
\affiliation{Department of Chemistry, Princeton University, Princeton, New Jersey 08544, USA}

\begin{abstract}
The broad success of theoretical and experimental quantum optimal control is intimately connected to the topology of the underlying control landscape.  For several common quantum control goals, including the maximization of an observable expectation value, the landscape has been shown to lack local optima if three assumptions are satisfied: (i) the quantum system is controllable, (ii) the Jacobian of the map from the control field to the evolution operator is full-rank, and (iii) the control field is not constrained. In the case of the observable objective, this favorable analysis shows that the associated landscape also contains saddles, i.e., critical points that are not local suboptimal extrema. In this paper, we investigate whether the presence of these saddles affects the trajectories of gradient-based searches for an optimal control. We show through simulations that both the detailed topology of the control landscape and the parameters of the system Hamiltonian influence whether the searches are attracted to a saddle. For some circumstances with a special initial state and target observable, optimizations may approach a saddle very closely, reducing the efficiency of the gradient algorithm. Encounters with such attractive saddles are found to be quite rare. Neither the presence of a large number of saddles on the control landscape nor a large number of system states increase the likelihood that a search will closely approach a saddle. Even for applications that encounter a saddle, well-designed gradient searches with carefully chosen algorithmic parameters will readily locate optimal controls.
\end{abstract}

\maketitle

\section{Introduction}
\label{sec:intro}

The last two decades have seen a significant expansion of the boundaries of quantum optimal control experiments (OCEs) due to technological advances in experimental resources, especially femtosecond lasers and pulse-shaping capabilities \cite{Rabitz2000, LevisRabitz2002, Goswami2003, BrixnerGerber2003, DantusLozovoy2004, Nuernberger2007, Brif2010NJP, WollenhauptBaumert2011, Brif2012ACP}. OCEs have been successfully performed for a wide range of goals, including the control of molecular vibrational \cite{HornungMeierMotzkus2000, WeinachtBartels2001CPL, BartelsWeinacht2002PRL, KonradiSingh2006JPPA, StrasfeldShim2007, KonradiScaria2007, ScariaKonradi2008, vanRhijnHerek2011} and electronic states \cite{BardeenYakovlevWilson1997, BrixnerDamrauer2001, Prokhorenko2005, Nahmias2005, BonacinaWolf2007, KurodaKleiman2009, vanderWalleHerek2009, RothGuyonRoslund2009, RoslundRothGuyon2011}, the generation and coherent manipulation of X-rays \cite{Bartels2000, Bartels2004, Reitze2004, PfeiferSpielmann2006, Winterfeldt2008}, the control of decoherence processes \cite{Branderhorst2008, BiercukBollinger2009}, the selective cleavage and formation of chemical bonds \cite{Assion1998, BergtBrixner1999, Levis2001, VajdaBartelt2001, Daniel2003, NuernbergerWolpert2010, Plenge2011}, the manipulation of energy flow in macromolecular complexes \cite{Herek2002, WohllebenBuckup2003, BuckupLebold2006, SavolainenHerek2008}, and the control of photoisomerization reactions \cite{VogtKrampert2005, VogtNuernberger2006CPL, DietzekYartsev2006, ProkhorenkoNagy2006, GreenfieldMcGrane2009}. Optimal control theory (OCT) \cite{RabitzZhu2000, DAlessandro2007, WerschnikGross2007, BalintKurti2008, Brif2010NJP, Brif2012ACP} has provided insights into the coherent control of a variety of quantum phenomena, such as electron transfer \cite{Kammerlander2011, Castro2012}, molecular photoisomerization \cite{OhtsukiOhara2003, ArtamonovHo2004CP, ArtamonovHo2006JCP, KurosakiArtamonov2009} and photodissociation \cite{Kosloff1989, ShiRabitz1991, Gross1992, NakagamiOhtsuki2002, Krieger2011}, the manipulation of trapped Bose-Einstein condensates \cite{HohenesterRekdal2007, HohenesterGrond2009, GrondHohenester2009}, strong-field ionization \cite{RasanenMadsen2012}, quantum information processing \cite{PalaoKosloff2002, TeschVivieRiedle2002, PalaoKosloff2003, KhanejaReiss2005, SchulteSporl2005, Hohenester2006, MontangeroCalarcoFazio2007PRL, GraceBrif2007JPB, VivieRiedleTroppmann2007, WeninPotz2008PRA, WeninPotz2008PRB, DominyRabitz2008JPA, Nebendahl2009, Schirmer2009JMO, SafaeiMontangero2009, WeninRoloffPotz2009, RoloffWeninPotz2009JCE, RoloffWeninPotz2009JCTN, Rebentrost2009PRL, RebentrostWilhelm2009, MotzoiGambetta2009PRL, GraceDominy2010NJP, SchulteHerbruggenSporl2011}, and spin squeezing in atomic ensembles \cite{TrailDeutsch2010, NorrisDeutsch2012}. 

The primary goal of OCEs and OCT simulations is to find a control $\varepsilon(t)$ that yields the global maximum or minimum value of a cost functional $J = J[\varepsilon(t)]$. This cost functional represents control objectives such as the distance between the unitary evolution operator and a target unitary transformation, the probability of a transition between two states, or the expectation value of an observable \cite{Brif2010NJP}. Several recent studies \cite{RabitzHsiehRosenthal2004, HoRabitz2006JPPA, MooreHsiehRabitz2008JCP} strongly indicate that the success of numerous OCEs and OCT simulations is related to the favorable topology of the quantum control landscape defined by the functional dependence of $J$ on $\varepsilon(t)$ \cite{Brif2010NJP, Brif2012ACP, ChakrabartiRabitz2007}.  In particular, it has been shown that the control landscape lacks local optima (referred to as traps) if three conditions are satisfied: (i) the quantum system is \textit{controllable}, i.e., any unitary evolution operator can be produced by some admissible control field beyond some finite time; (ii) the Jacobian matrix mapping the control field $\varepsilon(t)$ to the final-time evolution operator $U(T,0)$ is of full rank everywhere on the landscape; (iii) there are no constraints on the control field \cite{RabitzHsiehRosenthal2005PRA, HoRabitz2006JPPA, RabitzHoHsieh2006PRA, RabitzHsiehRosenthal2006JCP, HsiehRabitz2008PRA, WuRabitzHsieh2008JPA, HsiehWuRabitz2009JCP, HoDominyRabitz2009PRA, HsiehWuRabitzLidar2010}. The absence of local suboptimal extrema is of central importance to optimization; numerical studies have described the appearance of local traps on the control landscape due to the violation of assumption (i) \cite{WuHsiehRabitz2011PRA} and shown that the violation of assumption (ii) \cite{RivielloBrif2014} can, in special cases, prevent a gradient search from identifying globally optimal controls. Recent work \cite{RussellRabitz2016} has shown that assumptions (i) and (ii) are \emph{almost always} satisfied.  Thus, the satisfaction of assumption (iii) (which depends in practice on access to adequate system-specific control resources) is generally the key criterion that determines whether OCE or OCT searches will optimize successfully, especially with a local gradient-based algorithm. In this work, we assume that assumptions (i), (ii), and (iii) are satisfied and that the control landscape lacks local optima; this behavior was confirmed by the success of \textit{all} simulations.

Even when the three assumptions are satisfied, however, the control landscape for the unitary and observable objectives both contain sub-optimal critical points.  These critical points are saddles rather than local extrema, and cannot in principle trap a gradient-based search. However, gradient-based methods typically converge more slowly when they come near any such critical point. A prior numerical study of the unitary control objective indicated that saddles have little effect on gradient-based searches \cite{MooreChakrabarti2011}. In this work, therefore, we focus on observable control, for which the landscape may have a much larger number of saddles. Recent OCEs performed on a two-spin system located saddles on the observable control landscape at the predicted objective values and of the right character \cite{SunPelczer2015PRA}, providing empirical support for the theoretical analysis. 

The trajectory of a gradient search is influenced by both the landscape topology (which is fully defined by the initial state and target observable) and the local, non-topological geometry of the landscape (which depends on those two operators as well as the form of the Hamiltonian and the nature of the initial control field). We perform a large number of numerical OCT searches on a variety of control problems in order to identify physical parameters or characteristics that determine whether an optimal search will approach a saddle closely during an optimization. Using a specially designed metric \cite{SunRiviello2015}, we quantify the attractiveness of saddles and measure their influence on the efficiency of seeking optimal controls. The present work considers gradient-based simulations, which can be very sensitive to saddles.  In the laboratory, it is more common to employ stochastic algorithms, but the presence of a high density of attractive saddles could nonetheless be a challenge to optimization.  The findings in the present work are therefore relevant for effective performance in OCEs. 

The remainder of the paper is organized as follows: Section \ref{sec:backlandscape} discusses the theoretical basis for the classification of critical points, as well as the observable objective and the topology of the corresponding control landscape. Section \ref{sec:methodology} describes the numerical methods employed in this work and the metric used to evaluate the effects of saddles during a gradient-based search. In Section \ref{sec:results} we examine the factors that cause landscape saddles to influence searches for optimal controls.  Our concluding remarks are given in Section \ref{sec:conclusion}.

\section{Background and landscape analysis}
\label{sec:backlandscape}

\subsection{Background}
\label{sec:back}

The control illustrations in this paper involve closed $N$-level quantum systems with Hamiltonians of the form
\begin{equation}
\label{eq:ham}
H(t) = H_0 - \mu \varepsilon(t),
\end{equation}
within the electric dipole approximation. $H_0$ is the field-free diagonal operator, the control field $\varepsilon(t)$ is a real-valued function of time defined on the interval $[0,T]$, and $\mu$ is the dipole operator that couples the system to the field. In the Schr\"{o}dinger picture, the state of the system at a time $t$ is described by the density matrix $\rho(t) = U(t) \rho_0 U^{\dagger}(t)$, where $\rho_0 \equiv \rho(0)$ is the initial density matrix and $U(t) \equiv U(t,0)$ is the propagator or evolution operator. The propagator satisfies the Schr\"{o}dinger equation:
\begin{equation}
\label{eq:schro}
i \hbar \frac{d}{dt} U(t) = H(t) U(t) , \ \ \ U(0) = \mathbb{I} ,
\end{equation}
where $\mathbb{I}$ is the $N$-dimensional identity operator. In the present work, we only consider evolution-operator controllable systems \cite{DAlessandro2007, Brif2012ACP}; i.e., systems for which any unitary operator $W$ is the solution of the Schr\"{o}dinger equation (\ref{eq:schro}) at sufficently long time $T$ with some  control field $\varepsilon(t)$. In the absence of controllability, it has been shown that the control landscape may contain traps \cite{WuHsiehRabitz2011PRA}.

The topology of a quantum control landscape is determined by characterizing its \textit{critical points}, where
\begin{equation}
\label{eq:dyncrit}                                
\frac{\delta J}{\delta \varepsilon(t)} = 0 , \ \ \ \forall t \in [0,T] .
\end{equation}
Critical points can be classified as global extrema, local extrema, or saddles, according to the properties of second- and higher-order functional derivatives of $J$ with respect to the control field \cite{ChakrabartiRabitz2007, Brif2010NJP}. For example, the Hessian matrix,
\begin{equation*}
\label{eq:hess}
\mathsf{H}(t,t') = \frac{\delta^2 J}{\delta \varepsilon(t) \delta \varepsilon(t')} ,
\end{equation*}
describes the local curvature near a critical point. At a saddle, the Hessian has both positive and negative character. The existence of landscape saddles has practical significance for OCT optimizations, since their presence may influence searches with a gradient algorithm \cite{MooreRabitz2012, RivielloBrif2014} or even hinder the convergence efficiency of global stochastic algorithms \cite{DigalakisMargaritis2001}. The topic assessed in this paper is the role of saddles in seeking optimal controls, as reflected in the performance of a gradient-based algorithm which was chosen due to its sensitivity to landscape saddle features.

The landscape analysis for the objective $J$ can be performed using either the \emph{dynamic formulation}, in which the control landscape $J = J[\varepsilon(t)]$ is defined on the $L^2$ space of control fields, or the \textit{kinematic formulation}, in which the control landscape $J = J(U_T)$ is defined on the unitary group U$(N)$. In order to clarify the relationship between these two formulations, we partition the relationship between $J$ and the control field $\varepsilon(t)$ by representing $J$ as a function of the final-time evolution operator $U_T \equiv U(T)$, and $U_T$ in turn as a functional of the control field; i.e., $J = J(U_T)$ and $U_T = U_T [\varepsilon(t)]$. Using the chain rule, Eq.~\eqref{eq:dyncrit} can be rewritten as 
\begin{equation}
\label{eq:crit-1}
\frac{\delta J}{\delta \varepsilon(t)}
= \left\langle \nabla J(U_T), \frac{\delta U_T}{\delta \varepsilon(t)} \right\rangle = 0 ,
 \ \ \ \forall t \in [0,T] ,
\end{equation}
where $\langle \cdot , \cdot \rangle$ is the Hilbert-Schmidt inner product, $\nabla J(U_T)$ is the gradient of $J$ with respect to $U_T$, and the Jacobian matrix $\delta U_T / \delta \varepsilon(t)$ is the first-order functional derivative of $U_T$ with respect to the control field. Adopting satisfaction of assumption (ii), to which we referred in Sec.~\ref{sec:intro}, leads to the conclusion that Eq.~\eqref{eq:dyncrit} is equivalent to the kinematic result,
\begin{equation}
\label{eq:kincrit}
\nabla J(U_T) = 0 .
\end{equation} 
Therefore, the dynamic and kinematic perspectives yield the same landscape critical point specifications.

\subsection{Formulation and landscape topology of the control objective}
\label{sec:landscape}

The OCT simulations in this work consider the goal of maximizing the expectation value of a Hermitian quantum observable $\theta$ at time $T$:
\begin{equation}
\label{eq:tro}
J = \langle \theta(T) \rangle = \mathrm{Tr} ( U_T \rho_0 U_T^\dag \theta ). 
\end{equation}
In order to fully describe the landscape topology of this objective from the kinematic perspective $J(U_T)$, the multiplicities of the eigenvalues of $\rho_0$ and $\theta$ must also be specified.  Consider that $\rho_0$ has $r$ distinct eigenvalues $p_1 > p_2 > \ldots > p_r$ with corresponding multiplicities $a_1, a_2, \ldots , a_r$ and that $\theta$ has $q$ distinct eigenvalues $o_1 > o_2 > \ldots > o_q$ with corresponding multiplicities $b_1, b_2, \ldots , b_q$, where $q, r \leq N$.  It has been demonstrated that $\rho_0$ and $\theta$ can always be treated as diagonal in the eigenbasis of $H_0$ and with their eigenvalues sorted in descending order, i.e.,
\begin{align}
\begin{split}
\label{eq:eigsorted}
\rho_0 &= \text{diag} \{ p_1 , \ldots , p_1 ; \ldots ; p_r , \ldots p_r \} , \\
\theta &= \text{diag} \{ o_1 , \ldots , o_1 ; \ldots ; o_q , \ldots o_q \} ,
\end{split}
\end{align}
 with no loss of generality in the landscape analysis \cite{WuRabitzHsieh2008JPA}. In the kinematic formulation, it has also been shown that the sufficient and necessary condition for $U_T$ to be a critical point of the landscape is that the final-time density matrix $\rho(T) = U_T \rho_0 U_T^{\dag}$ commutes with the target observable $\theta$ \cite{HoRabitz2006JPPA, WuRabitzHsieh2008JPA, HsiehWuRabitz2009JCP}, i.e.,
\begin{equation}
\label{eq:trokincrit}
[\rho(T), \theta] = 0 .
\end{equation}
With $\rho_0$ and $\theta$ in the form of Eq.~\eqref{eq:eigsorted}, the condition in Eq.~\eqref{eq:trokincrit} is satisfied if and only if the unitary matrix $U_T$ lies in the double coset 
\begin{equation}
\label{eq:coset}
U_T = P \Pi Q^{\dag},  P \in \mathcal{U}(\mathbf{b}), Q \in \mathcal{U}(\mathbf{a})
\end{equation}
of some $N$-dimensional permutation matrix $\Pi$, where $\mathcal{U}(\mathbf{a}) = \mathcal{U}(a_1) \times \ldots \times \mathcal{U}(a_r)$ is the product of unitary groups of dimension $a_1, \ldots , a_r$ and $\mathcal{U}(\mathbf{b}) = \mathcal{U}(b_1) \times \ldots \times \mathcal{U}(b_q)$ is the product of unitary groups of dimension $b_1, \ldots , b_q$ \cite{WuRabitzHsieh2008JPA}.  In general, however, $\Pi$ is not unique and the evolution operators $U_T$ that satisfy Eq.~\eqref{eq:coset} are not permutation matrices.

By substituting Eq.~\eqref{eq:coset} into Eq.~\eqref{eq:tro}, the objective functional $J$ at a critical point can be rewritten as
\begin{equation}
\label{eq:eigperm}
J_{\text{crit}} = \mathrm{Tr} \left( P \Pi Q^{\dag} \rho_0 Q \Pi^{\dag} P^{\dag} \theta \right) =  \mathrm{Tr} \left( \Pi \rho_0 \Pi^{\dag} \theta \right).
\end{equation}
Thus, critical points on the observable objective landscape only exist at a finite number of discrete values of $J$; these values only depend on the eigenvalues of $\rho_0$ and $\theta$, not on the control field or Hamiltonian \cite{HoRabitz2006JPPA, WuRabitzHsieh2008JPA}. More specifically, each critical $J$ value corresponds to the sum of the product of the permuted eigenvalues of $\rho_0$ with the eigenvalues of $\theta$. Further characterization of the critical points of $J$ was accomplished via the \textit{contingency table} method described in \cite{WuRabitzHsieh2008JPA}. The contingency table $C$ is a $q \times r$ matrix whose nonnegative integer-valued elements $\{c_{jk}\}$, the so-called \textit{overlap numbers}, are the number of positions on the diagonals of $\theta$ and $\Pi \rho_0 \Pi^{\dag}$ where the distinct eigenvalues $o_j$ and $p_k$, respectively, both appear. The column and row sums of $C$ are $a_1, \ldots , a_r$ and $b_1, \ldots , b_q$, respectively.

A specific contingency table $C^i$ is shown in Table~\ref{tab:C}. The critical points of the landscape $J(U_T)$ that correspond to $C^i$ collectively comprise a \textit{critical submanifold} of the control landscape, which we denote as $M^i$.  All critical points in $M^i$ share the same objective value,
\begin{equation}
\label{eq:critvals}
J_i = \sum_{j,k=1}^{q,r} c_{jk}^i o_j p_k ,
\end{equation} 
although two critical submanifolds may have identical objective values.  We will denote the objective values corresponding to the global maximum and minimum of the landscape as $J_{\max}$ and $J_{\min}$, respectively. If both $\rho$ and $\theta$ are full rank, then each permutation $\Pi$ generates a distinct contingency table and thus there are $N!$ critical submanifolds on the landscape.  In this case, the critical submanifolds are disjoint $N$-tori, and analysis of the Hessian spectrum shows that two of them are the global maximum and global minimum of $J$ while the remainder are saddles \cite{HoRabitz2006JPPA, WuRabitzHsieh2008JPA}. Graphically, we can visualize the $i$-th critical submanifold as an infinitely thin ``pancake" of some shape in the function space of controls at its corresponding saddle value $J_i[\varepsilon(t)]$, where the gradient $\delta J_i / \delta \varepsilon(t) = 0$ and the Hessian $\mathsf{H}(t,t')$ has an indefinite non-zero spectrum and an infinite null space. If any eigenvalues of $\rho$ or $\theta$ are degenerate, then the same contingency table can be produced from multiple permutations $\Pi$, and the critical submanifold corresponding to that contingency table results from the merging of several $N$-tori. In this degenerate case, the landscape has fewer than $N! - 2$ saddles. The fewest landscape critical submanifolds arise when $\rho_0 = | i \rangle \langle i |$ and $\theta = | f \rangle \langle f |$, i.e., when $\rho_0$ and $\theta$ are projectors onto the pure states $| i \rangle$ and $| f \rangle$, respectively.  This special case of the observable objective is called the \textit{state-transition objective} and corresponds to maximizing the probability of a transition from $| i \rangle$ to $| f \rangle$.  The landscape for to state-transition control contains no saddles, so such problems are not considered in this paper; see Ref.~\cite{MooreRabitz2011} for a numerical study of state-transition landscapes.

\begin{table}[htbp]
\caption{\label{tab:C}The contingency table $C^i$, which describes an alignment between the distinct eigenvalues of $\rho_0$ and $\theta$  corresponding to the critical submanifold $M^i$.  The column and row sums of $C^i$ are $a_1, \ldots , a_r$ and $b_1, \ldots , b_q$, respectively.}
\begin{tabular}{c|ccc}
& $a_1$ & $\cdots$ & $a_r$ \\
\hline
$b_1$ & $c_{11}^i$ & $\cdots$ & $c_{1r}^i$ \\
$\vdots$ & $\vdots$ & $\ddots$ & $\vdots$ \\
$b_q$ & $c_{q1}^i$ & $\cdots$ & $c_{qr}^i$ \\
\end{tabular} 
\end{table}

For a particular $\rho_0$ and $\theta$, each permutation $\Pi$ leads to the construction of a (not necessarily unique) contingency table $C$, as described above. By repeating this process, all of the contingency tables for the landscape $J(U_T)$ can be determined, and the corresponding objective values indicate whether each table corresponds to the global maximum, the global minimum, or a saddle. As an example, consider a 4-level control problem with $\theta = \text{diag} \{ 0.5,0.2,0.2,0.1 \}$ and $\rho = \text{diag} \{ 0,0,0,1 \}$. $\theta$ has three distinct eigenvalues, $o_1 = 0.5$, $o_2 = 0.2$, and $o_3 = 0.1$, with multiplicities $b_1 = 1$, $b_2 = 2$, and $b_3 = 1$, respectively. $\rho_0$ has two distinct eigenvalues, $p_1 = 1$ and $p_2 = 0$, with multiplicities $a_1 = 1$ and $a_2 = 3$, respectively. Therefore, the contingency table corresponding to each critical submanifold is a $3 \times 2$ matrix with row sums  $\{ 1,2,1 \}$ and column sums $\{ 1,3 \}$. Under the permutation
\begin{equation*}
\renewcommand\arraystretch{0.75}
\Pi = \begin{pmatrix} 0 & 0 & 0 & 1 \\ 0 & 0 & 1 & 0 \\ 1 & 0 & 0 & 0 \\ 0 & 1 & 0 & 0 \end{pmatrix} ,
\end{equation*}
$\Pi \rho_0 \Pi^{\dag} = \text{diag} \{ 1,0,0,0 \}$. The overlap numbers for the contingency table corresponding to this permutation are determined by comparing the diagonal of the permuted density matrix with the diagonal of the observable $\theta$. $c_{11} = 1$ because the distinct eigenvalues $o_1 = 1$ and $p_1 = 0.5$ simultaneously appear at the first position (and no other positions) on the diagonals of $\theta$ and $\Pi \rho_0 \Pi^{\dag}$, respectively. Similarly, the remaining overlap numbers are determined to be $c_{21} = c_{31} = c_{12} = 0$, $c_{22} = 2$, and $c_{32} = 1$. If this process is repeated for all four-dimensional permutation matrices, then three distinct contingency tables are identified:
\begin{equation*}
\renewcommand\arraystretch{0.75}
C^1 = \begin{pmatrix} 1 & 0 \\ 0 & 2 \\ 0 & 1 \end{pmatrix} , \hspace{5mm} C^2 = \begin{pmatrix} 0 & 1 \\ 1 & 1 \\ 0 & 1 \end{pmatrix} , \hspace{5mm} C^3 = \begin{pmatrix} 0 & 1 \\ 0 & 2 \\ 1 & 0 \end{pmatrix} .
\end{equation*}
Using Eq.~\eqref{eq:critvals}, the objective values for each critical submanifold are calculated to be $J_1 = 0.5$, $J_2 = 0.2$, and $J_3 = 0.1$. Therefore, the contingency tables $C^1$ and $C^3$ correspond to the global maximum and minimum of the landscape, respectively, while $C^2$ corresponds to a saddle submanifold. The enumeration of these critical submanifolds fully describes the landscape topology for the observable control problem.

\section{Methodology}
\label{sec:methodology}

\subsection{Optimal control procedure}
\label{sec:contproc}

In this work, a gradient-based method will be employed to investigate local landscape saddle features because this procedure is ``myopic''; i.e., each step taken during the search is dictated by the local geometry of the control landscape at the current control field and thus is particularly sensitive to the presence of saddles. Each search is parameterized by the dimensionless index $s \geq 0$, which denotes the changes made to the field in the course of the optimization through the notation $\varepsilon(s,t)$.  The search trajectory is generated by solving the initial value problem
\begin{equation}
\label{eq:searchalg}
\frac{\partial \varepsilon(s,t)}{\partial s} = \gamma \frac{\delta J[\varepsilon(s,t)]}{\delta \varepsilon(s,t)} , 
\quad \varepsilon(0,t) \equiv \varepsilon_0(t) ,
\end{equation}
where the initial field is $\varepsilon_0(t)$, and the step size $\gamma$ is a positive constant. The functional derivative $\delta J / \delta \varepsilon(s,t)$ that appears in Eq.~\eqref{eq:searchalg} is calculated using the chain rule [as in Eq.~\eqref{eq:crit-1}] along with the previously-derived \cite{HoRabitz2006JPPA} relation 
\begin{equation*}
\label{eq:UT-deriv}
\frac{\delta U_T}{\delta \varepsilon(t)} = \frac{i}{\hbar} U_T U^{\dag}(t) \mu U(t) .
\end{equation*}
The result is \cite{HoRabitz2006JPPA, RabitzHoHsieh2006PRA, HoDominyRabitz2009PRA, MooreRabitz2012}:
\begin{equation}
\label{eq:gradJ}
\frac{\delta J}{\delta \varepsilon(t)} 
= \frac{2}{\hbar} \mathrm{Im} \mathrm{Tr} \left[ U_T^\dagger \theta U_T \rho_0 U^{\dag}(t) \mu U(t) \right] .
\end{equation}
We solve Eq.~\eqref{eq:searchalg} numerically using the MATLAB routine \texttt{ode45}, a fourth-order Runge-Kutta integrator with a variable step size (i.e., it determines $\gamma$ at each iteration) \cite{matlab}. \texttt{ode45} requires that an absolute error tolerance $\tau$ be specified, and we use the conservative value $\tau = 10^{-8}$ unless otherwise stated. The \textit{search effort}, defined as the number of iterations required for convergence, is an important measure of algorithmic efficiency. For the goal of maximizing the objective functional, the simulation is considered to have successfully converged when the search arrives at a control field $\varepsilon(s_f,t)$ that corresponds to an objective value $J \geq \left[ J_{\max} - 0.001 \cdot (J_{\max} - J_{\min}) \right ]$. 

In this paper, $\varepsilon(t)$ was discretized over $L$ evenly spaced intervals,
\begin{equation}
\label{eq:discrfield}
\varepsilon(t) = \{ \varepsilon_l | t \in (t_{l-1},t_l] \}_{l=1}^L , 
\end{equation}
where $t_l = l \Delta t$ and $\Delta t = T/L$. The overall evolution operator $U(t_l) \equiv U(t_l,0)$ is a product of incremental evolution operators,
\begin{subequations}
\begin{align*}
& U(t_l,t_{l-1}) = \exp \left[ -\frac{i}{\hbar} (H_0 - \mu \varepsilon_l) \Delta t \right] , \\
& U(t_l) = U(t_l,t_{l-1}) \cdots U(t_2,t_1) U(t_1,0) , 
\end{align*}
\end{subequations}
where the final-time evolution $U_T = U(t_L)$. The control variables are the $L$ real, independently-addressable field values $\{\varepsilon_l\}$, which can generate arbitrary pulse shapes as long as $L$ is sufficiently large. The $l$-th value of the initial field has the parameterized form 
\begin{equation}
\label{eq:field-init-1}
\varepsilon_l(0) \equiv \hspace{2mm} \varepsilon_0(t_l) = A(t_l) \sum_{m=1}^M a_m \cos (\omega_m t_l) , 
\end{equation}
where $A(t_l) = A_0 \exp \left[ -(t_l-T/2)^2 / (2 \eta^2) \right]$ is the Gaussian envelope function. The width of the envelope is specified by $\eta = T / 10$, and ensures that $\varepsilon_0(t) \approx 0$ at $t = 0$ and $t = T$. The $M = 20$ frequencies $\{ \omega_m \}$ are randomly selected from a uniform distribution on $[\omega_{\min},\omega_{\max}]$, where $\omega_{\min}$ and $\omega_{\max}$ are the smallest and largest transition frequencies in $H_0$, respectively. The amplitudes $\{ a_m \}$ are randomly selected from a uniform distribution on $[0, 1]$. The normalization constant $A_0$ is chosen so that the fluence, $F = \int_0^T \varepsilon^2(t) dt$, of the initial field $\varepsilon_0(t)$ has the value $F_0$.

After the field values $\varepsilon_l(s)$ are set in Eq.~\eqref{eq:field-init-1} (i.e., for $s > 0$), they are allowed to change according to the discrete version of Eq.~\eqref{eq:searchalg}:
\begin{equation}
\label{eq:searchalg-discr-1}
\frac{\partial \varepsilon_l(s)}{\partial s} = \gamma \frac{\delta J}{\delta \varepsilon_l(s)} \simeq \gamma \Delta t \frac{\partial J}{\partial \varepsilon(t_l)}. 
\end{equation}

\subsection{Critical distance metric}
\label{sec:metric}

The effect of saddles on a gradient search depends in part on how closely the search trajectory approaches the saddle submanifold.  We quantify this distance using the unitless \textit{critical distance metric} $D^i(U_T)$, which is a measure of the distance between a control $U_T$ and a particular critical submanifold $M^i$ on the kinematic observable landscape \cite{SunRiviello2015}.  Suppose that $\rho_0$ and $\theta$ are represented as diagonal matrices as in Eq.~\eqref{eq:eigsorted}, with their eigenvalues sorted in descending order.  $U_T$ can be divided into $q \times r$ rectangular blocks $U_{jk}$, each of dimension $b_j \times a_k$:
\begin{equation}
\label{eq:ublocks}
\renewcommand\arraystretch{0.75}
U_T = \begin{pmatrix} U_{11} & \cdots & U_{1r} \\ \vdots & \ddots & \vdots \\ U_{q1} & \cdots & U_{qr} \end{pmatrix}. 
\end{equation}
$U_{jk}$, which is generally not unitary, contains the elements of $U_T$ that correspond to the alignment of $o_j$ and $p_k$. Let the singular value decomposition of $U_{jk}$ be 
\begin{equation}
U_{jk} = X_{jk} S_{jk} Y_{jk}^{\dagger},
\end{equation}  
where the columns of the unitary matrices $X_{jk}$ and $Y_{jk}$ are the left and right singular vectors of $U_{jk}$, respectively, and $S_{jk}$ is a diagonal matrix containing the singular values $\sigma_{jkl}$ of $U_{jk}$ sorted in descending order. According to Theorem 1 of \cite{SunRiviello2015}, $U_T$ belongs to the critical submanifold $M^i$ of the observable landscape in Eq.~\eqref{eq:tro} if and only if the first $c_{jk}^i$ singular values $\sigma_{jkl}$ of each block $U_{jk}$ are equal to 1 and the remaining singular values of each block are equal to 0. Thus, the critical distance metric is defined by comparing each singular value of $U_{jk}$ to either 1 or 0, as appropriate \cite{SunRiviello2015}:
\begin{align}
\begin{split}
D^i(U_T) &= \sum_{j,k=1}^{q,r} \left [  \sum_{l \leq c_{jk}^i} \left ( 1 - \sigma_{jkl} \right )^2 + \sum_{l > c_{jk}^i} \sigma_{jkl}^2 \right ] \\
&= 2 \sum_{j,k=1}^{q,r} \sum_{l \leq c_{jk}^i} \left ( 1 - \sigma_{jkl} \right ) .
\end{split}
\end{align}
$D^i(U_T) = 0$ if and only if $U_T$ belongs to the critical submanifold $M^i$ corresponding to the contingency table $C^i$. For a particular control problem, the range of $D^i$ depends on the specific degeneracies of $\rho_0$ and $\theta$, and the maximum possible distance from each critical submanifold is not necessarily the same. For all control problems, however, the distance between any two critical submanifolds labeled $i$ and $i'$, defined as
\begin{equation}
\label{eq:twotabledist}
D^{i \to i'} = \sum\limits_{j,k=1}^{q,r} \left | c_{jk}^i - c_{jk}^{i'} \right | ,
\end{equation}
 cannot exceed $2N$. 

\section{Effect of saddles on gradient optimizations}
\label{sec:results}

Previous simulations of observable control problems with gradient-based algorithms have regularly reached the landscape maximum value \cite{RivielloBrif2014,RivielloMoore2015}. However, searches may converge more slowly while coming close to saddles, increasing the search effort. In this section, we investigate the practical effects of saddles on OCT simulations in a variety of control problems.  Many optimization parameters, details of the landscape topology, and features of the Hamiltonian affect whether searches approach saddles closely, so this paper cannot comprehensively address all the relevant aspects of any particular control problem. However, we discuss several key parameters that significantly influence saddle attraction.  Each parameter is studied, as independently of the others as possible, in order to evaluate its individual role. 

All simulations in this section are performed on one of two quantum systems, either rigid rotor-like,
\begin{equation}
\label{eq:rotor}
H_0 = \sum_{j=0}^{N-1} j(j+1) | j \rangle \langle j |,
\end{equation}
or an anharmonic oscillator,
\begin{flalign}
\label{eq:anharm}
H_0 &= \sum_{j=0}^{N-1} \left [ \kappa \left (j + \frac{1}{2} \right ) - \frac{\kappa^2}{\lambda} \left (j + \frac{1}{2} \right )^2 \right ] | j \rangle,&
\end{flalign}
where $\kappa = 2$ and $\lambda = 320$.  In both cases, the dipole matrix is 
\begin{equation}
\label{eq:dipole}
\mu = \sum_{j \neq k}^{N-1} \frac{d^{|j-k|}}{d} |j \rangle \langle k | ,	
\end{equation}
where the parameter $d \geq 0$. For the purposes of this paper, the field-free Hamiltonians $H_0$ in Eqs.~\eqref{eq:rotor} and \eqref{eq:anharm} were chosen to illustrate two extreme cases of increasing and decreasing energy level spacing, respectively. Correspondingly, the freedom in choosing $d$ in the dipole allows for sampling different degrees of coupling structure in $\mu$.

\subsection{Degeneracy of the initial state and the target observable}
\label{sec:degeneracy}

When the three assumptions described in Sec.~\ref{sec:intro} are satisfied, the topology of the observable landscape is fully determined by the number and multiplicities of the distinct eigenvalues of the initial density matrix $\rho_0$ and the target observable $\theta$.  Each permutation $\Pi$ of the eigenvalues of $\rho_0$ with respect to those of $\theta$ corresponds to a particular critical submanifold, as shown in Eq.~\eqref{eq:eigperm}. The multiplicities of the two sets of eigenvalues determine how many distinct permutations coincide with the global maximum, the global minimum, and each saddle \cite{WuRabitzHsieh2008JPA}.  If more than one permutation corresponds to a critical submanifold $M^i$, then evolution operators $U_T$ that coincide with $M^i$ can take a wider range of forms. Here, we perform optimizations on five related control problems in order to determine whether this additional freedom in the form of the critical $U_T$ influences the proximity of gradient searches to the saddles. Each search is performed on the rigid rotor-like system from Eqs.~\eqref{eq:rotor} and \eqref{eq:dipole}, with $N = 8$ and $d = 0.2$. The final time is $T = 20$ and the control period $[0,20]$ is divided into $L = 512$ intervals.  The initial state is $\rho_0 = | 0 \rangle \langle 0 |$; $\rho_0$ has two distinct eigenvalues $p_1 = 1$ and $p_0 = 0$, of multiplicities $a_1 = 1$ and $a_2 = 7$, respectively. Five target observables $\{\theta^m\}$ are considered:  
\begin{align}
\begin{split}
\label{eq:observ}
\theta^1 &= \frac{4}{9} | 6 \rangle \langle 6 | + \frac{5}{9} | 7 \rangle \langle 7 | , \\
\theta^2 &= \sum_{j=5}^6 \frac{4}{13} | j \rangle \langle j | + \frac{5}{13} | 7 \rangle \langle 7 | , \\
\theta^3 &= \sum_{j=4}^6 \frac{4}{17} | j \rangle \langle j | + \frac{5}{17} | 7 \rangle \langle 7 | , \\
\theta^4 &= \sum_{j=3}^6 \frac{4}{21} | j \rangle \langle j | + \frac{5}{21} | 7 \rangle \langle 7 | , \\
\theta^5 &= \sum_{j=2}^6 \frac{4}{25} | j \rangle \langle j | + \frac{5}{25} | 7 \rangle \langle 7 | . \\
\end{split}
\end{align}
 Each of the target observables $\theta^m$ have three distinct eigenvalues $o_1^m$, $o_2^m$, and $o_3^m$, with $o_3^m = 0$ in each case. The eigenvalues $o_1^m$ and $o_2^m$ are different for each control problem but chosen to ensure that every observable has unit trace: for example, $o_2^2 = 4/13$ and $o_1^2 = 5/13$.  In general, the multiplicities of the eigenvalues of $\theta^m$ are $b_1^m = 1$, $b_2^m = m$, and $b_3^m = 7-m$, respectively. The latter case of $b_3^m$ implies that the observable $\theta^m$ has $7-m$ zero eigenvalues, which are associated with the system states not explicitly shown in Eq.~\eqref{eq:observ}.

%$m$ denotes the mutltiplicity of the second largest eigenvalue of $\theta^m$.
 
As an example of using the method described in Sec.~\ref{sec:landscape}, the contingency tables for the control problem corresponding to the observable $\theta^5$ were determined to be
\begin{align}
\begin{split}
\label{eq:contintables}
\renewcommand\arraystretch{0.75}
C^{\max} &= \begin{pmatrix} 1 & 0 \\ 0 & 5 \\ 0 & 2 \end{pmatrix} , \\
C^{\text{sadd}} &= \begin{pmatrix} 0 & 1 \\ 1 & 4 \\ 0 & 2 \end{pmatrix} , \\ 
C^{\min} &= \begin{pmatrix} 0 & 1 \\ 0 & 5 \\ 1 & 1 \end{pmatrix} ,
\end{split}
\end{align}
with objective values $J_{\max} = 0.2$, $J_{\text{sadd}} = 0.16$, and $J_{\min} = 0$. Permutations that align the eigenvalue $p_1 = 1$ of $\rho_0$ with the eigenvalues $o_1^5 = 0.2$, $o_2^5 = 0.16$, and $o_3^5 = 0$ of $\theta^5$ correspond to the global maximum, saddle, and global minimum of the landscape, respectively. The contingency tables for the other four problems were constructed in the same way. Each control landscape contains one saddle, since every target observable has the same number of distinct nonzero eigenvalues and each control problem has the same initial state. In general, for each problem, permutations that align the eigenvalue $p_1$ with the eigenvalues $o_1^m$, $o_2^m$, and $o_3^m$ of $\theta^m$ correspond to the global maximum, saddle, and global minimum critical submanifolds, respectively. 

For each of the five control problems, we performed one hundred optimization runs using the control procedure described in Sec.~\ref{sec:contproc}. Each run began at a different initial field $\varepsilon_0(t)$ as defined in Eq.~\eqref{eq:field-init-1}, with fluence $F_0 = 10$. Every search converged successfully, and the results of these optimizations are reported in Table~\ref{tab:degen}. In addition, the distance to each of the critical submanifolds was calculated at every step of each optimization, using the distance metric described in Sec.~\ref{sec:metric}.  At an iteration of the gradient search denoted by the index $s$, the control field $\varepsilon(s,t)$ corresponds to an evolution operator $U_T(s)$ and in turn to a particular value $D^i[U_T(s)]$ of the critical distance metric for each critical submanifold $M^i$. The smallest value of $D^i$ over the interval $0 \leq s \leq s_f$ (i.e., the shortest distance to the saddle manifold $M^i$ at any point during a given search) was denoted as $D^i_{\min}$.  We use the mean value of $\overline{D^{\text{sadd}}_{\min}}$ for a set of one hundred optimizations to measure how closely gradient-based searches approach a saddle, on average, for a given control problem. For each set of optimizations, the mean search effort (MSE), i.e., the mean number of iterations, is also reported.

\begin{table}[htbp]
\caption{\label{tab:degen} Optimization results for several target observables $\theta^m$. The multiplicity of the second-largest eigenvalue $o_2^m$ of each observable is $m$. One hundred runs were performed for each observable.}
\begin{ruledtabular}
\begin{tabular}{cccc}
Observable & Multiplicity of $o_2^m$ & $\overline{D^{\text{sadd}}_{\min}}$ & MSE \\ \hline
$\theta^1$ & 1 & $3.35 \times 10^{-1}$ & 141 \\
$\theta^2$ & 2 & $2.34 \times 10^{-2}$ & 289 \\
$\theta^3$ & 3 & $8.12 \times 10^{-3}$ & 477 \\
$\theta^4$ & 4 & $7.36 \times 10^{-4}$ & 2745 \\
$\theta^5$ & 5 & $1.92 \times 10^{-4}$ & 9562 \\
\end{tabular}
\end{ruledtabular} 
\end{table}

\begin{figure}[htbp]
\centering
\includegraphics[width=8cm]{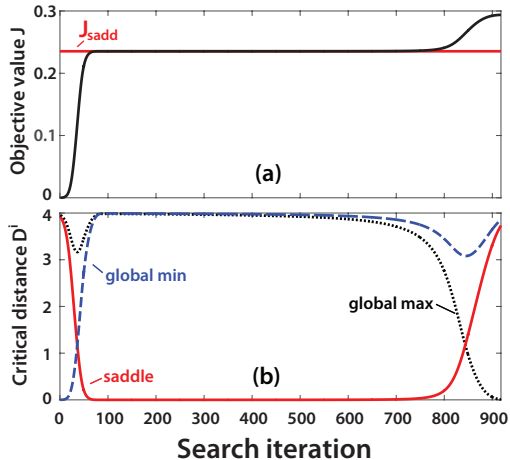}
\caption{(Color online) (a) The objective value at each iteration for a particular optimization run involving the target observable $\theta^3$ (black line) and the objective value of the saddle submanifold, $J_{\text{sadd}}$, (red line)  for this control problem.  (b) The value of the critical distance metric at each iteration of the same search, for $D^{\max}$ (dotted black line), $D^{\text{sadd}}$ (solid red line), and $D^{\min}$ (dashed blue line). \label{fig:degenrun}}
\end{figure}

These simulations indicate that gradient-based optimizations for which the multiplicity of the observable eigenvalue $o_2^m$ is larger involve a greater mean search effort and approach the saddle more closely. This trend is consistent with an expression for the dimension of critical submanifolds on the observable landscape obtained in Ref. \cite{WuRabitzHsieh2008JPA}.  Since the critical submanifold dimension does not take the system dynamics into account, its value is not predictive of the attractiveness of the saddle as measured in this paper.  However, the submanifold dimension qualitatively matches the trend in Table~\ref{tab:degen}; the dimension of the saddle increases with the multiplicity of $o_2^m$. Figure~\ref{fig:degenrun}, which illustrates the value of the objective $J$ and the distance to each critical submanifold at each iteration of a particular optimization run corresponding to $\theta^3$, shows that this increase in search effort results from a large number of iterations spent near the saddle submanifold.

Each of the control problems included in Table~\ref{tab:degen} has a control landscape with one saddle. The saddle corresponds to permutations $\Pi$ that align the eigenvalues $p_1$ (of $\rho_0$) and $o_2^m$ (of $\theta^m$). The correlation between the multiplicity of $o_2^m$ and the observed proximity to the saddle suggests that a broader range of critical unitary transformations $U_T$, made possible by this greater multiplicity, makes the saddle more attractive to a gradient search for these problems. In general, the simulations performed in this paper only encounter attractive saddles when $\rho_0$ and $\theta$ have respective eigenvalues that are highly degenerate and thereby have a strong influence on the nature of the saddle permutations $\Pi$. Importantly, we show later in Sec.~\ref{sec:numsaddles} that optimizations of several control problem cases with a large number of saddles have a \emph{significantly greater} mean distance of approach to any saddle than for the case with one saddle here. Thus, attractive saddles are expected to be rare in realistic control problems.

Other searches approached the saddle more closely than the example in Fig.~\ref{fig:degenrun}; one optimization corresponding to the observable $\theta^5$ required over $7 \times 10^4$ iterations to converge and reached a minimum distance of $D^{\text{sadd}}_{\min} = 8.95 \times 10^{-7}$ from the saddle. Despite the numerical challenges presented by such runs (for which the magnitude of the gradient becomes very small) none of the searches failed. These results corroborate prior numerical studies, which concluded that the observable objective is amenable to gradient-based optimization when the landscape lacks local extrema. Figure~\ref{fig:degenrun}(b) exhibits a notable feature of this optimization; after the search has passed its point of closest approach to the saddle submanifold and the objective value $J$ has passed $J_{\text{sadd}}$, the distance to the global minimum briefly decreases just as the gradient ascent resumes. In general, the non-monotonic behavior of the metric values $D^{\max}$, $D^{\text{sadd}}$, and $D^{\min}$ may reflect the varied and possibly complex shape of the critical submanifolds themselves over the space of controls. The phenomenon in Figure~\ref{fig:degenrun}(b), however, was observed in each of the optimizations summarized in Table~\ref{tab:degen} and is therefore unlikely to depend on the particular gradient path taken to the top of the landscape.  Instead, this behavior may be interpreted as a reflection of the relationship between the critical submanifolds for this set of control problems. Using Eq.~\eqref{eq:twotabledist}, the distance between \emph{any} two of the three critical submanifolds is calculated to be 4.  As a result of this symmetric relationship, each critical submanifold is at the maximum distance from the other two critical submanifolds, and thus any ascent or descent from one of them will be accompanied by an immediate decrease in the distance to all other critical submanifolds.

\subsection{Influence of Hamiltonian parameters}
\label{sec:hamiltonian}

The landscape topology, i.e., the characterization of the critical points of $J$, depends only on $\rho_0$ and $\theta$. Section~\ref{sec:degeneracy} shows that the nature of the topology has a significant effect on whether a gradient-based search is attracted to a saddle submanifold. However, the local geometry (i.e., the non-topological features) of the control landscape is also important and depends on the Hamiltonian.  For the Hamiltonian defined in Eq.~\eqref{eq:ham}, the dipole matrix $\mu$, the field-free Hamiltonian $H_0$, and the particular initial field $\varepsilon_0(t)$ each may influence whether an ascent of the landscape closely approaches saddles.  In this section, we independently consider the effect of each of these factors. 

The parameterization of the dipole matrix elements in Eq.~\eqref{eq:dipole} allows $\mu$ to take a variety of forms. For $d = 1$, all transitions $| j \rangle \to | k \rangle, j \neq k$ are equally allowed.  For $0 \leq d \leq 1$, all system transitions $| j \rangle \to | k \rangle, j \neq k$ are still allowed, but the value of the dipole moment coupling $\langle j | \mu | k \rangle$ decreases exponentially with the difference $|j-k|$. In the limit $d \to 0^+$, only transitions $| j \rangle \to | j \pm 1 \rangle$ between adjacent system states are allowed. To determine the effect of the dipole coupling parameter $d$ on the attraction of a gradient search to the saddle, we performed a set of optimizations utilizing the same $\rho_0 = | 0 \rangle \langle 0|$ and $\theta^m$ that were used in Sec.~\ref{sec:degeneracy}. In particular, consider the target observable $\theta^5 = \sum_{j=2}^6 0.16 | j \rangle \langle j | + 0.2 | 7 \rangle \langle 7 |$, which we will denote as Case (I) in the remainder of this work. The dipole matrix is given in Eq.~\eqref{eq:dipole}, and simulations were performed using both the rigid rotor-like and anharmonic oscillator field-free Hamiltonians [Eqs.~\eqref{eq:rotor} and \eqref{eq:anharm}, respectively]. In all simulations, the system had $N = 8$ levels, the control interval of $T = 20$ was divided into $L = 512$ steps, and the initial field fluence was $F_0 = 10$. The contingency tables corresponding to Case (I) are given in Eq.~\eqref{eq:contintables}. Accordingly, the landscape contains three critical submanifolds: the global maximum, the global minimum, and a saddle.  The high search effort associated with Case (I) (i.e., the observable $\theta^5$) for the rotor-like Hamiltonian in Table~\ref{tab:degen} indicates that the saddle is very attractive to gradient-based searches when the dipole parameter $d = 0.2$ is used.  In this section, we selected values of the dipole parameter over the range $0.2 \leq d \leq 1$ and we performed one hundred optimizations for each value of $d$ using both forms of $H_0$. The distance to each critical submanifold was measured using the metric $D^i$, and the mean of the smallest distance to the saddle, $\overline{D^{\text{sadd}}_{\min}}$, was determined for each set of one hundred runs.

\begin{figure}[htbp]
\centering
\includegraphics[width=8cm]{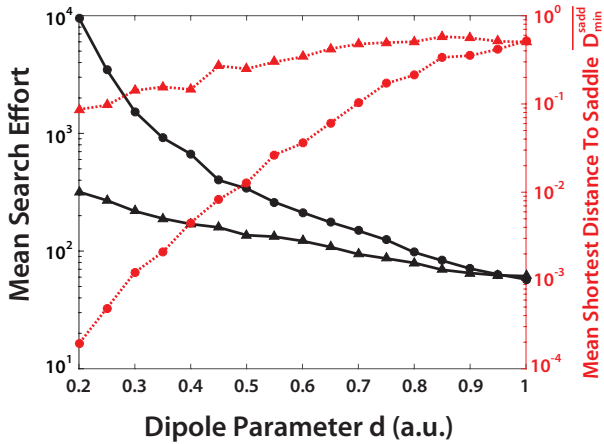}
\caption{(Color online) The mean search effort (solid black lines) and the mean shortest distance to the saddle submanifold $\overline{D^{\text{sadd}}_{\min}}$ (dotted red lines), as a function of the dipole parameter $d$, for optimizations of Case (I) that use the rigid rotor-like (circles) and anharmonic oscillator (triangles) systems. Both systems have $N = 8$ levels, and one hundred optimization runs were performed for each value of $d$. \label{fig:Ddrop}}
\end{figure}

Figure~\ref{fig:Ddrop} shows that for both choices of $H_0$, larger values of $d$ correspond to smaller search effort and to searches that are less attracted to the saddle. The dipole moment for transitions between adjacent states is the same for any value of $d$; i.e, $\langle j \pm 1 | \mu | j \rangle = 1$ $\forall d$.  However, the dipole moment for a transition between non-adjacent states decreases as $d$ decreases. Therefore, these results indicate that the landscape saddle for this control problem is more attractive to gradient searches when non-adjacent states are coupled less strongly to one another. While the control problem corresponding to the observable $\theta^5$ contains a very attractive saddle for $d = 0.2$, as observed in Sec.~\ref{sec:degeneracy}, searches performed for larger values of $d$ do not approach the saddle closely. This result demonstrates that the values of the dipole elements play a significant role in determining the effect of saddles on gradient optimizations of Case (I).

The expression for the objective $J$ at critical points in Eq.~\eqref{eq:eigperm} helps to clarify the role of the dipole matrix elements in this control problem. For the simulations in this section, the global maximum of the landscape corresponds to evolution operators $U_T$ that align $p_1 = 1$ and $o_1^5 = 0.2$, which are the largest eigenvalues of $\rho_0$ and $\theta^5$, respectively. For Case (I), the density matrix eigenvalues satisfy $\rho_j = 0 \forall j \geq 2$, so aligning $p_1$ and $o_1^5$ strictly assures an optimal solution at the top of the landscape that lies on the global maximum submanifold of optimal solutions. As Eq.~\eqref{eq:ublocks} shows, the alignment of any pair of eigenvalues from the initial state and the target observable, respectively, corresponds to one of the blocks $U_{jk}$ into which we divide $U_T$.  Since both $p_1$ and $o_1^5$ are of unit multiplicity, the block $U_{11}$ is a single element of $U_T$. The relevant element is $\langle 7 | U_T | 0 \rangle$, since $p_1$ is associated with the ground state $| 0 \rangle$ of the system and $o_1^5$ is associated with the state $| 7 \rangle$. Therefore, the optimal objective value is achieved via an alignment of $\rho_0$ and $\theta$ that corresponds to the system transition $|0 \rangle \to | 7 \rangle$.

For $d \approx 1$, the dipole moment $\langle 7 | \mu | 0 \rangle = d^6$ for this transition may be large enough to achieve the alignment of $\rho_0$ and $\theta$ required for a globally optimal control directly, via a $|0 \rangle \to | 7 \rangle$ transition. In contrast, for small values of $d$, the dipole moment for the $|0 \rangle \to | 7 \rangle$ transition is much smaller than the dipole moment for transitions between adjacent states.  As a result, the optimal evolution operator $U_T$ is more likely to correspond to a series of adjacent-state transitions (e.g. $|0 \rangle \to |1 \rangle \to \ldots \to | 7 \rangle$) constructively interfering along with additional companion pathways in order to reach the top of the landscape. Analogously, the saddle submanifold corresponds to unitary transformations that align the eigenvalue $p_1 = 1$ with the eigenvalue $o_2^5 = 0.16$.  The expression for $\theta^5$ in Eq.~\eqref{eq:observ} shows that $o_2^5$ encompasses the states $| 2 \rangle, | 3 \rangle, \ldots | 6 \rangle$, so the transitions $| 0 \rangle \to | j \rangle, 2 \leq j \leq 6$ are associated with the saddle.  The dipole moment for each of these transitions is larger than the dipole moment for the transition $|0 \rangle \to | 7 \rangle$ associated with the global maximum. For Case (I), this disparity grows as $d$ decreases, making it more likely that the gradient search will come even closer to the saddle.    

Figure~\ref{fig:Ddrop} also indicates that the relationship between the dipole parameter $d$ and the attractiveness of the saddle is much more dramatic for the rigid rotor-like system than for the anharmonic oscillator system. While the two field-free Hamiltonians lead to similar values of the mean search effort and of $\overline{D^{\text{sadd}}_{\min}}$ when $d = 1$, smaller values of the dipole parameter lead to a disparity between the rotor-like and oscillator optimizations. When $d = 0.2$, the rigid rotor-like simulations required a mean search effort of 9562 iterations and led to a mean shortest saddle distance of $\overline{D^{\text{sadd}}_{\min}} = 1.92 \times 10^{-4}$. On the other hand, the anharmonic oscillator simulations required a mean search effort of 318 iterations and a mean shortest saddle distance of $\overline{D^{\text{sadd}}_{\min}} = 8.66 \times 10^{-2}$. Therefore, the form of the field-free Hamiltonian also influences the attractiveness of the saddle in Case (I). However, the correlation between the dipole parameter $d$ and the attractiveness the saddle is qualitatively similar for both the rotor-like and oscillator systems.  The same trends were observed with optimizations using other observables $\theta^m$ (not shown here) for the anharmonic oscillator system; as in the rigid rotor simulations summarized in Table~\ref{tab:degen}, a greater multiplicity of the second-largest observable eigenvalue corresponded to searches that approached the saddle more closely.

Additionally, the strength of the field-system interaction is proportional to both the transition dipole moment and the amplitude of the control field. Therefore, a transition for which the dipole moment is very small can still occur with significant probability if the amplitude of the field is sufficiently large.  To determine whether the amplitude of the control field $\varepsilon(t)$ influences the form taken by $U_T$ during the course of an optimization and thus affects whether gradient optimizations are attracted to the saddle, we performed additional sets of simulations with with larger values of the initial field fluence $F_0$ over the range $50 \leq F_0 \leq 1000$. These optimizations used the rigid rotor-like system in Eqs.~\eqref{eq:rotor} and \eqref{eq:dipole}, with $d = 0.2$.  All other parameters were the same as for the prior simulations in this section. One hundred optimizations were performed for each value of $F_0$.

\begin{table}[htbp]
\caption{\label{tab:fluence} Optimization results for Case (I), for various values of the initial field fluence $F_0$. The rigid rotor-like system with $N = 8$ states were used, and the dipole parameter $d = 0.2$. One hundred optimization runs were performed for each $F_0$ value.}
\begin{ruledtabular}
\begin{tabular}{rcc}
$F_0$ & $\overline{D^{\text{sadd}}_{\min}}$ & MSE \\ \hline
$10^1$ & $1.92 \times 10^{-4}$ & 9562 \\
$5 \times 10^1$ & $2.66 \times 10^{-2}$ & 572 \\
$10^2$ & $1.13 \times 10^{-1}$ & 372 \\
$5 \times 10^2$ & $4.57 \times 10^{-1}$ & 179 \\
$10^3$ & $4.89 \times 10^{-1}$ & 150 \\
\end{tabular}
\end{ruledtabular} 
\end{table}

As reported in Table~\ref{tab:fluence}, the use of a larger initial field fluence significantly reduced both the search effort and the mean proximity to the saddle during a search. Comparing this result to Fig.~\ref{fig:Ddrop}, we find that gradient optimizations of Case (I) are significantly attracted to the saddle submanifold only when both the dipole parameter $d$ and the intial field fluence $F_0$ are sufficiently small. These results emphasize the importance of carefully choosing optimization parameters, such as the initial fluence, in order for an OCT search of this problem to be as efficient as possible. Even if the critical topology of the landscape and the form of the dipole matrix yield saddles that are likely to attract a gradient search, an optimization of Case (I) is unlikely to approach these saddles if the amplitude of the field is sufficiently large. The collective results from Table~\ref{tab:fluence} reflect that the rapidly evolving search at high initial fluence quickly passes the saddle, in contrast to the behavior in Fig.~\ref{fig:degenrun}.

\subsection{The number of saddles}
\label{sec:numsaddles}

The optimizations in Secs.~\ref{sec:degeneracy} -- \ref{sec:hamiltonian} were performed on 8-level control problems for which the landscape has only one saddle submanifold, due to the degeneracy of the initial state $\rho_0$ (i.e., all but one of the eigenvalues were equal to zero) and the nature of the target observable. Thus, we showed that saddles may attract gradient searches when $\rho_0$ is a projector onto a pure state and $\theta$ has the particular structure described in Eq.~\eqref{eq:observ}.  In contrast, the landscape can contain a greater number of saddles when $\rho_0$ and $\theta$ have different structures, reaching the extreme when both operators are full rank and nondegenerate. In this section, we investigate the effect of a large number of saddles by performing optimizations of $J$ on two additional control problem cases:
\begin{enumerate}[(I)]
\setcounter{enumi}{1}
\item The initial density matrix is $\rho_0 = | 0 \rangle \langle 0 |$, as in Case (I), but a different full rank target observable $\theta$ is used for each simulation. Each observable is defined as 
\begin{equation}
\label{eq:fullrankO}
\theta = \mathcal{N}_o \sum_{j=1}^N \tilde{o}_j | j \rangle \langle j | ,
\end{equation} 
where the normalization $\mathcal{N}_o = 1 / \sum\limits_{j} \tilde{o}_j$, and each of the values $\tilde{o}_j$ are randomly selected from a uniform distribution on the interval $[0,1]$. The critical topology for this problem was determined using the methods in Sec.~\ref{sec:landscape}. There are eight contingency tables, each of which is an $8 \times 2$ matrix with column sums $a_1 = 1, a_2 = 7$ and row sums $b_k = 1 \hspace{2mm} \forall k$:
\begin{equation}
\renewcommand\arraystretch{0.75}
C^1 = \begin{pmatrix} 1 & 0 \\ 0 & 1 \\ 0 & 1 \\0 & 1 \\0 & 1 \\0 & 1 \\0 & 1 \\0 & 1 \end{pmatrix}, \hspace{5mm} \ldots \hspace{5mm} , C^8 = \begin{pmatrix} 0 & 1 \\ 0 & 1 \\ 0 & 1 \\0 & 1 \\0 & 1 \\0 & 1 \\0 & 1 \\1 & 0 \end{pmatrix} .
\end{equation}
$C^1$ corresponds to the global maximum, $C^8$ corresponds to the global minimum, and the remaining contingency tables correspond to the six saddle submanifolds of the landscape.  For this Case (II), the expression in Eq.~\eqref{eq:critvals} for the critical objective values simplifies to $J_i = o_i$.
\item The initial density matrix and the target observable are both full rank and lack degeneracy, and each optimization uses a different choice of both $\rho_0$ and $\theta$. The observable is defined as in Eq.~\eqref{eq:fullrankO}, and $\rho_0$ is analogously defined as 
\begin{equation}
\label{eq:fullrankeigs}
\rho_0 = \mathcal{N}_p \sum_{k=1}^N \tilde{p}_k | k \rangle \langle k |,
\end{equation}
where the normalization $\mathcal{N}_p = 1 / \sum\limits_{k} \tilde{p}_k$ and each value $\tilde{p}_k$ is selected randomly from the interval $[0,1]$. Each of the $8! = 40320$ contingency tables is one of the $N$-dimensional permutation matrices $\Pi$ and its associated critical objective value is determined using Eq.~\eqref{eq:eigperm}. Two of the critical submanifolds correspond to the global maximum and minimum, and the remainder are saddles.
\end{enumerate}

Cases (II) and (III) differ from each other only in the form of $\rho_0$. For both cases, the rigid rotor-like system from Eqs.~\eqref{eq:rotor} and \eqref{eq:dipole} is used, with $N = 8$ levels and the dipole parameter is $d = 0.2$.  The final time is $T = 20$ and the control period is divided into $L = 512$ intervals. A random initial control field $\varepsilon_0(t)$ with a fluence $F_0 = 10$ was used for each search, and one hundred runs were performed for Cases (II) and (III). The distance $D^i(U_T)$ to each critical submanifold was mesasured during every run. The control landscape for Cases (II) and (III) have many saddles, and $D^{\text{sadd}}_{\min}$ is defined as the shortest distance to \emph{any} saddle during an optimization. The results of these optimizations are reported in Table~\ref{tab:numsadd}, with the results for Case (I) (from Sec.~\ref{sec:degeneracy}) included for comparison.

\begin{table}[htbp]
\caption{\label{tab:numsadd} Optimization results for Cases (I) -- (III). For each case, the rigid rotor-like system with $N = 8$ levels and dipole parameter $d = 0.2$ was used. One hundred optimization runs were performed for each control case.}
\begin{ruledtabular}
\begin{tabular}{cccccc}
Case & $\rho_0$ & $\theta$ & No. of saddles & $\overline{D^{\text{sadd}}_{\min}}$ & MSE \\ \hline
(I) & Pure state & $\theta^5$ & 1 & $1.92 \times 10^{-4}$ & 9562 \\
(II) & Pure state & Full rank & 6 & $9.93 \times 10^{-2}$ & 302 \\
(III) & Full rank & Full rank & 40318 & $5.56 \times 10^{-1}$ & 332 \\
\end{tabular}
\end{ruledtabular} 
\end{table}

\begin{figure}[htbp]
\centering
\includegraphics[width=8cm]{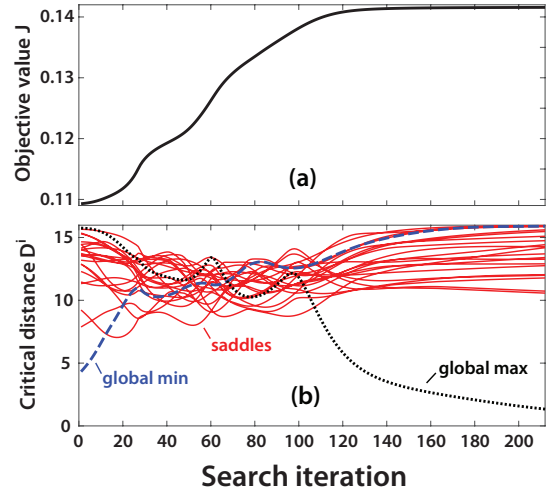}
\caption{(Color online) (a) The objective value at each iteration for an optimization of Case (III). (b) The value of the critical distance metric at each iteration of the same search, for $D^{\max}$ (dotted black line), $D^{\min}$ (dashed blue line), and $D^{\text{sadd}}$ for twenty randomly-selected saddles (solid red lines). \label{fig:fullrankrun}}
\end{figure}

For Case (I), the single saddle on the control landscape was extremely attractive to a gradient search.  However, when the same parameters were used for Cases (II) and (III), the optimizations did not approach any saddle very closely, especially for Case (III). For Case (II), the mean shortest distance to any saddle was $\overline{D^{\text{sadd}}_{\min}} = 9.93 \times 10^{-2}$, and no search passed closer than $D^{\text{sadd}}_{\min} = 4.45 \times 10^{-3}$ to any saddle. For Case (III), the mean shortest distance to any saddle was $\overline{D^{\text{sadd}}_{\min}} = 0.556$, and no search passed closer than $D^{\text{sadd}}_{\min} = 0.194$ to any saddle. Thus, the landscape saddles for Cases (II) and (III) are much less likely to attract gradient searches than the saddle in Case (I). Figure~\ref{fig:fullrankrun} demonstrates this point by illustrating the value of the objective $J$ and the distance to twenty randomly-selected critical submanifolds at each iteration of one optimization of Case (III). The large number of saddles on the corresponding control landscape made it necessary to use a random sample for graphical purposes, but this sample is qualitatively representative of the entire set of 40320 critical submanifolds. The single saddle for the optimization in Fig.~\ref{fig:degenrun} is far more attractive than any of the saddles represented in Fig.~\ref{fig:fullrankrun}.  While the small ``kinks'' in Fig.~\ref{fig:fullrankrun}(a) indicate points at which the optimization was attracted to a saddle, the distance to each of the $8! - 2 = 40318$ saddles was monitored for each optimization and none of them exhibit more than these minor effects.

For Cases (II) and (III), the eigenvalues of $\rho_0$ and $\theta$ that must be aligned in order for a control to lie on the saddle all have unit multiplicity. Therefore, the simulations in this section support our previous conclusion that the multiplicity of the eigenvalues is correlated with attractive landscape saddles for these cases. Furthermore, this result suggests that when the landscape has a large number of saddle submanifolds, a gradient search is less likely to to be significantly attracted to any one of them. This conclusion is significant for the control of nominally complex systems where $\rho_0$ and $\theta$ may have high rank.

\subsection{The number of system states}
\label{sec:numlevels}

For the optimizations in Secs.~\ref{sec:degeneracy} -- \ref{sec:hamiltonian},  the control landscape has only one saddle submanifold due to the degeneracy in the initial state and the observable.  For the optimizations in Sec.~\ref{sec:numsaddles}, one or both of $\theta$ and $\rho_0$ are full rank and the resulting control landscape has many saddles. In addition, all of the previous simulations in this paper were performed on eight-level systems. Since problems of physical interest often involve systems with many states, in this section we investigate whether the effect of saddles on a gradient search depends on the number of levels $N$.

All simulations in this section were performed on the rigid rotor-like system from Eqs.~\eqref{eq:rotor} and \eqref{eq:dipole}. Gradient searches were performed for control Cases (I) -- (III), and each case was generalized to $N$-level systems.  For Case (I), the initial state and observable are still defined as in Sec.~\ref{sec:hamiltonian}, with $\rho_0 = | 0 \rangle \langle 0 |$ and $\theta^5 = \sum_{j=2}^6 0.16 | j \rangle \langle j | + 0.2 | 7 \rangle \langle 7 |$. Therefore, the only change to the eigenvalue spectra of $\rho_0$ and $\theta$ for different values of $N$ is the multiplicity of the smallest (zero) eigenvalue of each operator. For all values of $N$, the landscape contains three critical submanifolds, which again correspond to the global maximum, the global minimum, and a single saddle. For Case (II), there are $N$ critical submanifolds, of which $N - 2$ are saddles.  The contingency tables are the set of $N \times 2$ matrices with column sums $a_1 = 1, a_2 = N-1$ and row sums $b_k = 1, 1 \leq k \leq N$.  For Case (III), there are $N!$ critical submanifolds, of which $N! - 2$ are saddles. Each contingency table is one of the $N$-dimensional permutation matrices $\Pi$.

\begin{figure}[htbp]
\centering
\includegraphics[width=8cm]{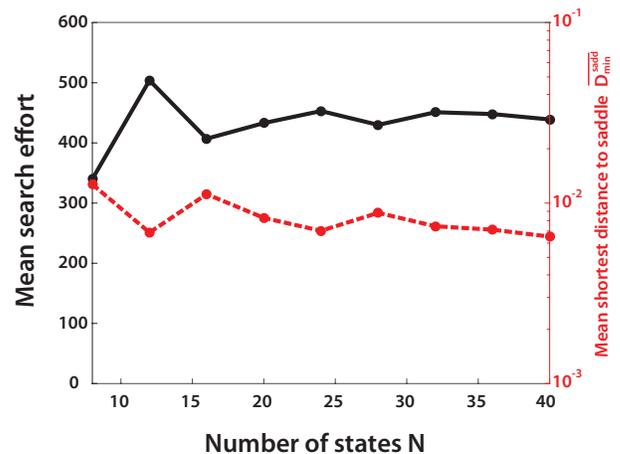}
\caption{(Color online) The mean search effort (solid black line) and the mean shortest distance to the saddle submanifold, $\overline{D^{\text{sadd}}_{\min}}$ (dashed red line) as a function of the number of states $N$, for gradient optimizations of Case (I). The dipole parameter $d = 0.5$, and one hundred optimization runs were performed for each value of $N$. \label{fig:degenscale}}
\end{figure}

First, we performed optimizations of Case (I) for a number of states ranging over $12 \leq N \leq 40$.  The dipole parameter was $d = 0.5$, the final time was $T = 20$, and $L = 512$ time intervals were used.  One hundred optimizations using initial fields with fluence $F_0 = 10$ were performed for each value of $N$, and the distance $D^i$ to the saddle submanifold was monitored during each optimization.  The results are illustrated in Fig.~\ref{fig:degenscale}, with the prior results for $N = 8$ and $d = 0.5$ from Sec.~\ref{sec:hamiltonian} included as well. They indicate that both the mean search effort and the mean value of $D^{\text{sadd}}_{\min}$ for Case (I) remain relatively constant regardless of the number of states. As discussed in Sec.~\ref{sec:degeneracy}, the saddle submanifold for this case corresponds to controls that align the eigenvalues $p_1$ and $o_2$ of $\rho_0$ and $\theta$, respectively.  Since the multiplicity of neither of these eigenvalues increases with $N$, it is intuitive that the attractiveness of the saddle is relatively invariant to the number of system states. In addition, all system states $| j \rangle, j > 7$ are associated with the smallest (i.e., zero) eigenvalue of both the initial state and the target observable. Evidently, the higher-lying states play a very limited role in the optimal search.

Optimizations were also performed for Cases (II) and (III), for $3 \leq N \leq 16$ (the simulations in Sec.~\ref{sec:numsaddles} used $N = 8$). In these simulations, the dipole parameter $d = 0.2$ and the control time $T = 20$ was divided into $L = 512$ intervals for $N < 10$ and into $L = 2048$ intervals for $N \geq 10$. The initial field fluence was $F_0 = 10$, and one hundred optimizations were performed for each value of $N$. The distance $D^i$ to the saddle submanifold was measured at each step of the search for every value of $N$ for Case (II) and for $3 \leq N \leq 10$ for Case (III). These values were used to determine $D^{\text{sadd}}_{\min}$, which is defined as in Sec.~\ref{sec:numsaddles}, i.e., the shortest distance to any saddle during a given optimization. For Cases (II) and (III), we also define the quantity $D^{\text{mean(sadd)}}_{\min}$ as the mean of the shortest distance to each saddle for a given optimization. Neither measure was calculated for $N > 10$ in Case (III) due to the factorial scaling of the number of critical submanifolds (e.g., for $N = 11$, the landscape contains $\sim 3.99 \times 10^7$ saddles).

\begin{figure}[htbp]
\centering
\includegraphics[width=8cm]{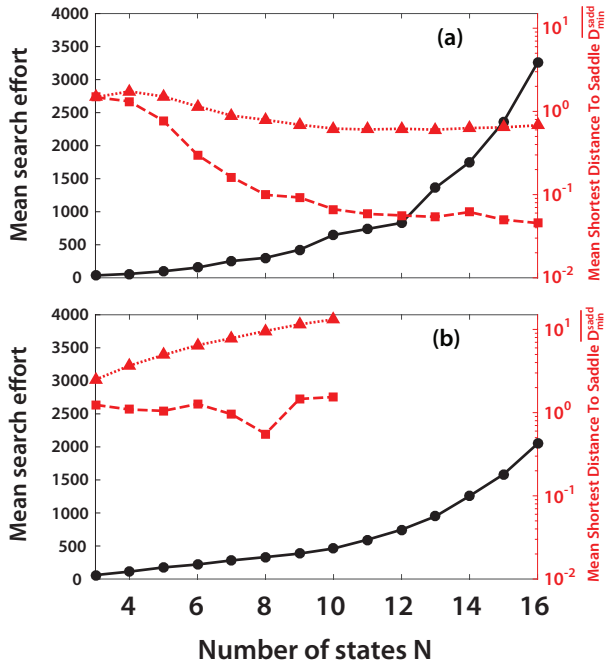}
\caption{(Color online) The mean search effort (black circles), the mean shortest distance to any saddle submanifold, $\overline{D^{\text{sadd}}_{\min}}$ (red squares), and the mean shortest distance averaged over all saddle submanifolds, $\overline{D^{\text{mean(sadd)}}_{\min}}$ (red triangles) as a function of the number of states $N$, for gradient optimizations of (a) Case (II) and (b) Case (III). The dipole parameter $d = 0.2$, and one hundred optimization runs were performed for each value of $N$. \label{fig:fullrankscale_d2}}
\end{figure}

Figure~\ref{fig:fullrankscale_d2} illustrates that the search effort increases with the number of system states $N$ for both cases, reflecting the complexity of these control problems in many-level systems. For Case (I), a critical unitary evolution $U_T$ is optimal (i.e., corresponds to the global maximum of the landscape) if and only if it aligns the largest eigenvalue of $\rho_0$ with the largest eigenvalue of $\theta$.  For Case (III), however, a critical $U_T$ must simultaneously aligns each of the $N$ eigenvalues of the initial state $\rho_0$ with a particular eigenvalue of the target observable $\theta$ in order to be optimal. Despite this scaling with $N$, the mean search effort for Cases (II) and (III) at $N = 16$ (3255 and 2049 iterations, respectively)  was still significantly less than for optimizations of Case (I) for $N = 8$ and $d = 0.2$ (9562 iterations). Once again, we observe that additional system complexity does not significantly impact the effort of optimization for these cases, even with a greater number of saddles present. Amongst these various cases, there also exist other subtle trends that must result from the details of the dynamics involved. 

Additionally, Fig.~\ref{fig:fullrankscale_d2} shows that the mean value $\overline{D^{\text{sadd}}_{\min}}$ remains comfortably large for all values of $N$ for Cases (II) and (III), indicating that optimizations do not closely approach any saddles. For Case (II), the mean shortest distance to any saddle was $\overline{D^{\text{sadd}}_{\min}} = 1.48$ for $N = 3$; while this value initially decreases as $N$ grows, it remains relatively constant at $\overline{D^{\text{sadd}}_{\min}} \approx 0.08$ for $N \geq 8$. Most saddles are not approached even this closely, as the mean shortest distance averaged over all saddles was $\overline{D^{\text{mean(sadd)}}_{\min}} \approx 0.7$ for $N \geq 8$. For Case (III), $\overline{D^{\text{sadd}}_{\min}} \approx 1$ for all values of $N$, while $\overline{D^{\text{mean(sadd)}}_{\min}} = 2.5$ for $N = 3$ and increases to $\overline{D^{\text{mean(sadd)}}_{\min}} = 13.2$ for $N = 10$. Although the number of saddles on the landscape corresponding to Case (III) increases factorially with $N$, the trend in $\overline{D^{\text{mean(sadd)}}_{\min}}$ indicates that the average attractiveness of each saddle decreases with $N$. This behavior is consistent with a mathematical analysis of the kinematic volume fraction near critical submanifolds that was performed on a related observable problem \cite{DominyRabitz2011JPA}. These competing trends may explain why, although the number of saddles increases by a factor of $3.5 \times 10^{12}$ between $N = 3$ and $N = 16$, the mean search effort only increases by a factor of 34 over the same range (60 for $N = 3$ and 2049 for $N = 16$). This dramatic disparity shows that Case (III) is surprisingly amenable to gradient optimization, despite its very large number of landscape saddles.

\begin{figure}[htbp]
\centering
\includegraphics[width=8cm]{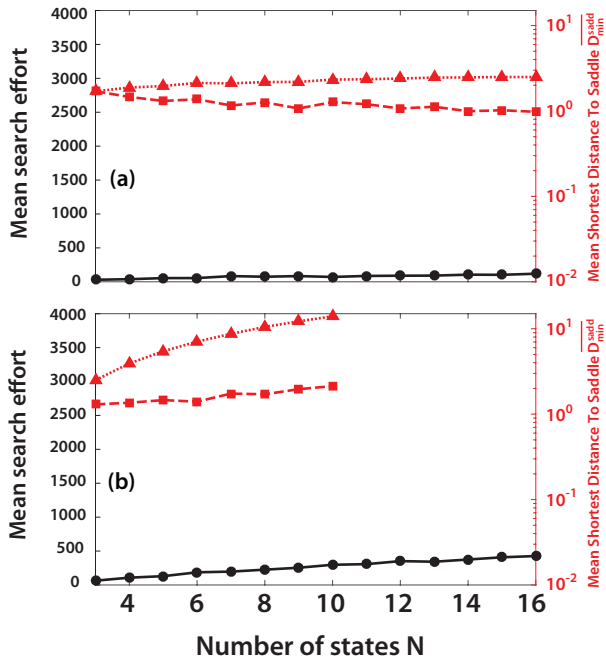}
\caption{(Color online) The mean search effort (black circles), the mean shortest distance to any saddle submanifold, $\overline{D^{\text{sadd}}_{\min}}$ (red squares), and the mean shortest distance averaged over all saddle submanifolds, $\overline{D^{\text{mean(sadd)}}_{\min}}$ (red triangles) as a function of the number of states $N$, for gradient optimizations of (a) Case (II) and (b) Case (III). The dipole parameter $d = 1$, and one hundred optimization runs were performed for each value of $N$. \label{fig:fullrankscale_d10}}
\end{figure}

Like the saddle attraction measured in Sec.~\ref{sec:hamiltonian}, the observed scaling of search effort with $N$ for Cases (II) and (III) also depends on Hamiltonian parameters.  We repeated the optimizations represented in Fig.~\ref{fig:fullrankscale_d2} in the same manner as described above, but with the dipole parameter $d = 1$ rather than $d = 0.2$.  The results of this set of simulations are illustrated in Fig.~\ref{fig:fullrankscale_d10}.  All measures of the distance to saddles are at least equal to the values observed for $d = 0.2$, and the search effort scaling is significantly less; at $N = 16$, the mean effort for Cases (II) and (III) is 119 and 427, respectively. The strong coupling may have accelerated the searches for these cases by preventing them from lingering near any of the saddles.

In conclusion, the optimizations of Case (I) in this section show that increasing the number of system states does not affect the search effort or the attractiveness of the saddle submanifold.  While the mean search effort increases with $N$ for Cases (II) and (III), we demonstrate that this scaling is not due the saddles becoming more attractive, as the point of closest approach to a saddle does not change significantly with $N$.  On average, the degree of attraction to any individual saddle appears invariant to $N$ for Case (II) and decreases with $N$ for Case (III).

\subsection{Control constraints}
\label{sec:constraints}

In the \texttt{ode45} algorithm, the accuracy demanded of the solutions to Eq.~\eqref{eq:searchalg} is determined by the error tolerance $\tau$. In a previous numerical study that used a gradient algorithm to investigate the role of control constraints in OCT optimization, the effect of changing this tolerance was studied \cite{RivielloMoore2015} and it was determined that a choice of $\tau = 10^{-8}$ yields sufficiently accurate solutions to find optimal controls for the state-transition objective. It was also shown that large values of $\tau$, and the resulting inaccurate solutions to Eq.~\eqref{eq:searchalg}, are a severe constraint that may result in search failure (i.e., a decrease in the value of the objective functional after an iteration). However, the simulations in \cite{RivielloMoore2015} involved the state-transition landscape, which lacks saddles. When a gradient search is close to a critical point of the landscape, the norm of the gradient is small, and more accuracy may be required of the solutions to Eq.~\eqref{eq:searchalg} in order to ensure successful optimization. Therefore, a smaller value of $\tau$ may need to be used when the control landscape has saddles, particularly if the saddles attract gradient searches. The simulations in this section explore whether less accurate solutions to Eq.~\eqref{eq:searchalg} can cause searches to fail in close proximity to a saddle submanifold.

We performed optimizations of Case (I) as defined in Sec.~\ref{sec:hamiltonian}, with $N = 8$ levels and $d = 0.2$. Case (I) was chosen because its control landscape contains the most attractive saddle identified in this work. The control period was $T = 20$, the time discretization was $L = 512$, and each initial field had fluence $F_0 = 10$. One hundred runs were performed for each value of the absolute error tolerance $\tau$ over the range $10^{-7} \leq \tau \leq 10^{-1}$. For each failed optimization, the distance to the saddle at the final iteration, $D^{\text{sadd}}_{\text{fail}}$, was recorded. The results of these optimizations are reported in Table~\ref{tab:tol}, and they confirm that a large value of $\tau$ will cause searches that use \texttt{ode45} to fail.  All searches failed for $\tau \geq 10^{-2}$, and at least one search failed for $\tau \geq 10^{-6}$. The mean distance from the saddle at which searches fail increases with $\tau$, suggesting that the appropriate error tolerance for a particular optimization of this problem is determined by the attractiveness of the saddle(s) on the corresponding control landscape. When the search approaches a saddle more closely, a smaller value of $\tau$ is required in order to avoid search failure. The results of the simulations in Ref.~\cite{RivielloMoore2015} support this conclusion; for a control problem that lacks saddles, all optimizations were successful for $\tau \leq 2 \times 10^{-3}$.  The searches on a landscape that has an attractive saddle require significantly more accurate solutions to Eq.~\eqref{eq:searchalg}, as $\tau \leq 10^{-7}$ is required to ensure that all searches succeed. Thus, the choice of $\tau = 10^{-8}$ for this paper is adequate for the field to reach its optimal form.

\begin{table}[htbp]
\caption{\label{tab:tol} Optimization results for Case (I) and various values of $\tau$, the absolute error tolerance in \texttt{ode45}. The system has $N = 8$ states and one hundred optimization runs were performed for each $\tau$ value.}
\begin{ruledtabular}
\begin{tabular}{ccl}
$\tau$ & No. failed & $\overline{D^{\text{sadd}}_{\text{fail}}}$ \\ \hline
$10^{-1}$ & 100 & 1.66 \\
$10^{-2}$ & 100 & $ 1.35 \times 10^{-2}$ \\
$10^{-3}$ & 97 & $ 2.68 \times 10^{-4}$ \\
$10^{-4}$ & 89 & $ 9.20 \times 10^{-5}$ \\
$10^{-5}$ & 50 & $ 1.41 \times 10^{-5}$ \\
$10^{-6}$ & 14 & $ 2.10 \times 10^{-6}$ \\
$10^{-7}$ & 0 & - \\
\end{tabular}
\end{ruledtabular} 
\end{table}

\section{Conclusion}
\label{sec:conclusion}

The critical topology of the quantum control landscape has been analyzed theoretically \cite{HoRabitz2006JPPA, WuRabitzHsieh2008JPA, Altafini2009, ChakrabartiRabitz2007, WuPechenRabitz2008JMP, RabitzHsiehRosenthal2004, RabitzHoHsieh2006PRA, RabitzHsiehRosenthal2005PRA, DominyRabitz2008JPA, HsiehHoRabitz2008CP, ShenHsiehRabitz2006JCP} in consideration of the mounting successes of diverse optimal control experiments and simulations. These theoretical works have shown that the landscape lacks local optima when three assumptions are met: controllability, a full-rank Jacobian matrix $\delta U_T / \delta \varepsilon(t)$ everywhere on the landscape, and an unconstrained control field $\varepsilon(t)$.  A recent numerical work \cite{RivielloMoore2015} suggests that only the latter condition is of prime importance to avoid significant resource constraints. Satisfaction of these assumptions ensures that all intermediate critical points (i.e., those that do not correspond to the global maximum or minimum) are saddles. For the observable objective considered here, the initial state $\rho_0$ and the target observable $\theta$ determine the topology of the control landscape, which may have as many as $N! -2$ saddle submanifolds for an $N$-level system. This paper has investigated the effect of these saddles on gradient searches. 

At one extreme, we identified a control problem, Case (I), with a landscape that contains a highly attractive saddle submanifold (i.e., a saddle that almost all gradient searches will approach closely at some point during the optimization). For this problem, the majority of algorithmic iterations takes place very close to the saddle; the distance to the saddle was calculated using the critical distance metric \cite{SunRiviello2015}. We also identified features of the landscape topology that lead to this phenomenon. For this case, a saddle submanifold is more attractive to a gradient search when the eigenvalues that correspond to the saddle alignment have greater multiplicity, and the proximity of the search to the saddle is also influenced by parameters of the system Hamiltonian; optimizations experience a significantly greater attraction to the saddle when the field-free Hamiltonian has a rigid rotor-like energy structure rather than that of an anharmonic oscillator. Even with a rigid rotor-like $H_0$ and the particular degeneracy of $\rho_0$ and $\theta$ described above, the saddle in Case (I) is only observed to attract gradient searches when the dipole moment for transitions between non-adjacent states and the initial fluence of the control field are both sufficiently small. This choice of values may result in a tendency for the gradient search to initially drive the unitary evolution operator towards alignments of the eigenvalues of $\rho_0$ and $\theta$ that correspond to a saddle submanifold before optimizing, rather than driving the evolution toward the global maximum directly. 

We also studied cases for which the target observable, or both the initial state and the observable, are full rank.  The landscapes for these problems have multiple saddles (a very large number, in the latter case). Importantly, we demonstrated that the greater number of saddles for these cases does not imply a correspondingly greater probability that a search will closely approach \emph{any} saddle. In the case of the control problem for which the landscape contains the largest possible number of saddles, the average attractiveness of the saddles \emph{decreases} significantly as $N$ \emph{increases}. This result has significance for many normally complex laboratory circumstances where a high density of occupied states may be involved and the observable also involves many states. In this regard as well, the gradient-based algorithm used in this work is likely the most conservative method; typical use of stochastic search algorithms in the laboratory likely will be less sensitive to saddles, as they may be able to ``step over" them. Thousands of numerical OCT searches were performed in this paper, and they were only significantly attracted to a saddle for a very particular choice of the initial state, target observable, and several Hamiltonian parameters.  The great majority of the optimizations that were performed do not approach any saddle closely. Even for a control problem that corresponds to a landscape with an attractive saddle, it was shown that a careful choice of algorithmic parameters ensures successful optimization. This conclusion is based on the extensive numerical studies performed for several control problem cases in this work. While there is still the need for a rigorous mathematical understanding of its origin that builds on the foundations in Ref. \cite{DominyRabitz2011JPA}, the results for these cases indicated that control landscapes free of local optima are highly favorable for optimization, even when saddles are present.

\acknowledgments

G.R. acknowledges support from the Department of Energy under Grant No. DE-FG02-02ER15344, and R.B.W. acknowledges support from the National Natural Science Foundation of China. Q.S. acknowledges support from the Army Research Office under Grant No. W911NF-16-1-0014 and from the Princeton Plasma Science and Technology Program. H.R. acknowledges support from the National Science Foundation under Grant No. CHE-1464569.

\bibliographystyle{apsrev4-1}
\bibliography{riviellorabitz_saddles}

\end{document}